# Single Crystal Growth Tricks and Treats

Tanya Berry[†,‡,§], Nicholas Ng[†,‡], Tyrel M. McQueen[†,‡,¶]

† Department of Chemistry, The Johns Hopkins University, Baltimore, Maryland 21218, United States

‡ Institute for Quantum Matter, William H. Miller III Department of Physics and Astronomy, The Johns Hopkins University, Baltimore, Maryland 21218, United States

§ Department of Chemistry, PrincetonUniversity, Princeton, NewJersey 08544, UnitedStates

¶ Department of Materials Science and Engineering, The Johns Hopkins University, Baltimore, Maryland 21218, United States

Corresponding email: tberry@ucdavis.edu, nng3@jhu.edu, and mcqueen@jhu.edu

Conspectus Single crystal growth is a widely explored method of synthesizing materials in the solid state. The last few decades have seen significant improvements in the techniques used to synthesize single crystals, but there has been comparatively little discussion on ways to disseminate this knowledge. We aim to change that. Here we describe the principles of known single crystal growth techniques as well as lesser-known variations that have assisted in the optimization of defect control in known materials. We offer a perspective on how to think about these synthesis methods in a grand scheme. We consider the temperature interdependence with the reaction time as well as ways to carry out synthesis to scale up and address some outstanding synthesis challenges. We hope our descriptions will aid in technological advancements as well as further developments to gain even better control over synthesis.

Introduction Human civilization has always needed useful materials in order to advance, grow, and support the demands of society. This necessity has proven true since the first stone tools were made, and improvements have been constantly sought up to the present day. Materials are needed for most facets of modern life today, from energy production, transmission and storage to refrigeration, transportation and modern medicine. Because of this need, the search both for new materials and new ways of making these new materials has been a constantly active area of research.[1-3]

The traditionally taught lore in materials is that the physical properties follow from the crystal structure of the material. However, the exact mechanisms by which certain structural motifs give rise to desirable properties are of significant interest to our field of study. Studying these structural motifs helps us better understand both how current materials exhibit desirable phenomena as well as provides insight into structural motifs that we can then exploit in the design of new and improved materials.[4-24] It is also useful for attempting to create new materials that exhibit emergent properties that are of interest to cutting-edge technologies. Quantum materials are one area of extensive interest.[25-27] This term describes materials whose physical properties cannot be sufficiently described by classical physics and includes materials such as superconductors, magnetic, and topological materials.[28-43] These materials are important to materials science because they represent a likely next generation of materials that can augment and improve upon existing technologies, as well as potentially enable entirely new ones.[44-46]

Realizing useful emergent quantum phenomena such as superconductivity, magnetism, and topology in materials is key to advancing both fundamental understanding and technological advancements.[44-51] Some significant materials that exhibit superconductivity include yttrium barium copper oxide, or BSCCO,[52-54] NbTi alloys,[55] and $Nb_3Sn$.[55] These materials are all superconducting materials, some of which are currently in commercial use today. In terms of magnetism, understanding the diversity of ground states and spin behavior is an area of rich interest both in discovering quantum spin liquid materials as well as in technological applications such as spintronics, particularly with the discovery of materials such as but not limited to $RuCl_3$[56], herbertsmithite[57], $Fe_3Sn_2$[58-59], $KV_3Sb_5$[60-61], $Mn_3Ge$[62], yttrium aluminum garnet[63-64], yttrium iron garnet[65], $CrO_2$[66-67], Ni-Co alloys[68-71], $Mn_3Sn$[72] and even elements such as platinum[73-74], among others[75-76]. that have brought new areas of study and application within magnetism. It is crucial to study the nature of axis-dependent magnetism in order to learn about spin orientations. Topology also offers research and technological richness, from the discovery of topological insulators and Dirac semi-metals in materials to probing applications in thermoelectricity and catalysis.[46,77] Materials such as $Bi_2Te_3$,[78] $Bi_2Se_3$,[79] $Na_3Bi$,[80] and $Cd_3As_2$[81] have allowed us to better understand topology by realizing properties such as the quantum Hall effect, spin Hall effect, and other transport properties that have yielded electrical conductivity ideal for various technological applications. As noted in magnetism, topological materials are sometimes also anisotropic, and studying the nature of transport requires the community to study these materials either in the thin film or single crystal forms.

Single crystals are also crucial to materials discovery. They typically have low defect concentrations overall and because they possess only a single crystalline domain, physical properties are consistent throughout the whole crystal, which can be important both for getting a true picture of what the material is capable of as well as being the easiest form to put into use commercially. They are also generally high purity materials, which are also very useful for sensitive applications where small impurity concentrations can cause problems.

The large time gap between the discovery of the material and the ability to easily manufacture it on an industrial scale illustrates one of the primary challenges of materials discovery—optimizing the process in order to minimize the time from fundamental research to commercialization[82-84]. The act of discovering these materials is itself a relatively difficult process and one potential way to make it easier is to understand the mechanisms underlying these desirable physical phenomena. As such, exploring available avenues and optimizing both known and new material synthesis becomes extremely important.

Here, readers may ask why it is important to optimize materials, especially ones that are already commercialized and in regular use in technology worldwide. A major reason is to enable defect control to allow new applications. Defects (which can also include, e.g., variations in elemental isotopes) in crystalline materials are ubiquitous; it is possible to minimize them with very fine control over the synthetic process, as can be done with silicon, but entropy demands that they be present. Thus, since it is not entirely possible to grow the perfect single crystal, it is important to make these defects work to our advantage[85-90]. Defect control can influence physical properties both by tuning desired properties and by introducing new physical properties.[91-95]

Furthermore, the precise control of synthetic reactions is necessary in order to target new materials that can be very difficult to make. However, the knowledge of the systematic effects of different parameters on the reaction and the final crystalline product is often not known prior to conducting the experiment. For example, the reaction temperature is one of the easiest parameters to effectively control, yet optimal temperature ranges for synthesizing as-yet-unknown materials are difficult to determine without performing experiments. Applying high pressure to materials during or after synthesis can influence the crystal structure, and by extension the physical properties of the material as well[96-97]. Additionally, the choice of crystallization technique can also influence successful materials discovery. Selecting a technique to turn material from either powder or other parent reagents is nontrivial and requires careful consideration of the many factors involved[98-108]. However, the properties of many materials that influence their suitability for certain techniques are not known and can be difficult to measure. For example, crystallizing a material by laser floating zone requires that the target material absorb the monochromatic laser light, but in many cases, the optical absorption spectrum of the target will not be known until the experiment is already in progress.

The field of materials science demands a wide variety of synthesis methods that are suitable for growing single-domain crystals in order to discover emergent physical phenomena. Crystallization is to some extent a very well-studied field[100, 105-106, 109-131]—most are familiar with the traditional solvent-based recrystallization methods—but in the area of condensed matter materials synthesis, many more useful routes to single crystals exist that are not as well known. The lack of knowledge for many techniques, such as floating zones, microwave synthesis, and spark plasma sintering, prevents scientists from accessing the full range of synthetic methods that are available. Many of these less well-known approaches also possess variants on the original procedure which provide additional flexibility. However, because these techniques are not as widely used within the community, the necessary considerations for selecting one to use for material synthesis are also less well-known. In this paper, an overview of many condensed matter materials synthesis techniques will be presented alongside potential sticking points that need to be considered prior to selecting one to use. Variations and deviations from the typical classes of materials synthesized by certain techniques will also be discussed. Herein, we draw out solid state synthesis tricks that may not be traditional and hope to offer synthetic strategies. In the pursuit of new materials, it is imperative to look towards areas that are not well explored.

## Materials & Methods

There are a wide variety of material classes and a similar range of available methods for high-quality crystal growth. Available methods include but are not limited to the flux growth, chemical vapor transport, Czochralski and Bridgman techniques, spark plasma sintering, hydrothermal/solvothermal synthesis, and floating zone growth, and are summarized in Figure 1. Different methods are useful for growing different materials, examples of which are shown in Table 1.

The flux growth method is like the traditional wet chemistry solution crystal growth in that materials are combined in a container, typically an alumina crucible, and dissolved to facilitate the reaction[126-131]. The flux, however, is typically some kind of solid that has a low melting point instead of a liquid solvent and is added in excess to provide sufficient dissolving power. As in traditional solution chemistry, the most important aspect of the chosen flux is that it should be able to dissolve all the reagents in order to facilitate the reaction between them. After the reaction is complete, the flux is typically slowly cooled to just above its melting point, then centrifuged in order to remove the excess flux. There are a wide variety of materials used as fluxes for this type of crystal growth, and many different materials have been crystallized using this technique[132-134]. Examples of crystalline materials grown via the flux method include oxides[127,135-138], sulfides[139-140], antimonides[61,141], as well as nitrides such as hexagonal BN[142] and GaN[143]. Important considerations include the choice of crucible material (it must be unreactive toward both the flux and dissolved species), and the binary/ternary/quaternary phase diagrams which dictate under what conditions the liquid is in equilibrium with the desired solid product.

Arc melting uses a high-voltage generator to generate a current through a conductive metal line that ends in an electrode, typically made from thoriated tungsten. This electrode is sharpened to a fine point and when close to or in contact with a conducting surface, it can generate an electric arc, which is used as a direct heat source to melt materials such as refractory elements or compounds together to achieve reaction[144-146]. Many materials targeted for synthesis via arc melting are some forms of intermetallic compounds, such as medium[147]- and high-entropy[148-149] alloys, as well as simpler intermetallics such as TiAl[150], AlCo[151], and CoSn[152]. However, arc melting has also been used to make other materials such as $CeAlO_3$/$CeGaO_3$[153], BN nanotubes[154] and a superconducting sulfide, $Zr_{3+x}S_4$[155].

As the name states, microwave crystal growth uses microwave radiation to heat materials in order to facilitate the reaction. It is possible to generate heating in this manner either by direct irradiation of the target materials or by secondary heating, where the microwaves irradiate and heat up a secondary bystander material that then radiates heat to the target materials[156-157]. This secondary heating method is very useful if the target materials do not absorb microwaves well, but results in reduced heating power. However, overall microwave heating accelerates the crystal growth process to the point where the material forms in the order of minutes, making it a fast and valuable option for growing crystals of certain materials, especially those containing nearly unstable or volatile constituents. Microwave heating has been in explicit use since at least 1986 as an organic synthesis technique[158], but recently has seen more application towards condensed matter. Some examples of materials grown by microwave heating include ZnS nanoballs[159], metal-organic frameworks[160], and several 2D layered metal oxides[161], Ge nanocrystals[162], and $SnO_2$ nanocrystals[163-164] as well as a number of doped metal oxides that are used as phosphor materials in LEDs[165-166].

Hydrothermal synthesis is different from almost all other solid state crystal growth techniques in that it takes place in a solution that is chemically very different from the species being grown- water[167]. The selection of different temperature and pressure conditions can greatly impact the species formed in solution and thus can influence the reaction outcome. At a high enough temperature and pressure, the reaction can be brought to a supercritical state which changes the properties of the solvent[168-169], allowing for reactions that might not otherwise occur. These conditions also allow to some degree the use of materials that may otherwise have low solubility in the chosen solvent to grow crystals. Hydrothermal synthesis, like many of the other techniques described here, has been used to synthesize a large variety of materials[170-172] such as the halide perovskite $CsPbBr_3$[173], GaN[174], CdS[175], as well as a number of different transition metal oxides[125, 176-179]. Hydrothermal synthesis has also been combined with microwave heating to synthesize Eu-doped $SrTiO_3$[180].

A more general form of hydrothermal synthesis is to replace water with non-aqueous solvents – so-called solvothermal synthesis. As with traditional wet chemistry methods, the choice of solvent is extremely important and can significantly affect the production of the target material. Often used now in the preparation of metal-organic framework single crystals[181-183], the method has a long history[184-185] and has been used to carry out novel solid state chemical transformations, such as in the preparation of formally $Ni^{1+}$ phases (e.g. $La_4Ni_3O_8$[186-187]). Another approach is to use an electrochemical force to drive formation of the desired phase[188-189]. Long used for production of electroactive elements such as aluminum (the Hall-Héroult process[190-191]), it has also been used to great effect in the preparation of novel molecular-based quantum phases[192]. A key strength of electrochemical methods is that they allow precise control over the thermodynamic driving force of formation, and enable creation of oxidation states that are otherwise broadly inaccessible in single crystalline form. A limitation is that it works best for materials that are electrically conductive and able to support appropriate transmission of the electrochemical potential to the growth interface.

Chemical vapor transport (CVT) is a form of crystal growth that generates crystals via gas phase transport. The target material is combined with a volatile transport agent in a long, sealed tube under vacuum. When this tube is heated, the transport agent is vaporized. The transport agent then reacts with the target material, transforming it into a gaseous intermediate phase which can then travel along the length of the tube. Once the intermediate reaches the other, typically cooler, end of the tube, the temperature is low enough that there is no longer sufficient thermal energy to maintain the gaseous intermediate. It then decomposes into the transport agent, which remains in the gas phase, and the target material, which crystallizes and deposits on the tube surfaces[193]. A wide variety of materials have been synthesized using CVT reactions, and the technique has been extensively researched in order to analyze the different factors that go into a successful application of the method[113, 194-196], particularly the crucial choice of a transport agent[114]. In particular, they are very useful at growing 2D layered materials[104, 115, 197] in addition to other compounds such as zinc oxide[198]. Variants of CVT combine with other techniques – e.g. one can use a salt solution in place of a gaseous reagent, making a version that combines aspects of hydrothermal and CVT growth methods. Non-standard containers for CVT also enable greater flexibility -- e.g. light sealed alumina/zirconia tubes to access higher temperatures, or sealed platinum or gold containers. CVT, being a very slow and typically layer-by-layer growth process, tends to yield very high quality, unstrained crystal specimens.

Bridgman crystal growth moves the target material through a static hot zone, with crystallization occurring at a single point where the temperature gradient is designed to be particularly sharp to traverse from a few degrees above to a few degrees below the melting point[116-117]. The rate of movement is typically very slow, in the order of millimeters per hour, in order to facilitate slow recrystallization and obtain the highest quality single crystal possible. There are multiple methods of generating the hot zone; one method that we will briefly focus on here is the induction method, which is useful when the target material is metallic or semiconducting. A large electric current is run through coils of conducting wire, which induces a current through the target material. The material then heats up and melts via resistive heating. In addition, Bridgman growth can take advantage of zone refinement purification, as impurities in the bulk material tend to stay in the liquid phase. Some of the first materials grown using this method include elemental crystals by Bridgman[116] and lithium fluoride by Stockbarger[117]. Recently, this technique has become primarily useful for slow, large crystal growth of both intermetallics[199] and oxides[63], among other compounds.

Czochralski growth is a method famously used for growing extremely large boules of single crystal semiconductor elements such as silicon[118] and germanium[119-120]. For the typical growth, a large amount of the target material is completely melted in preparation. A small single crystal seed is lowered into the melt, then pulled out at a very slow rate. The molten material crystallizes as it is removed from the heat, ideally taking on the same crystallographic characteristics as the seed that was used to begin the growth and forming a very large single crystalline domain[200]. This technique has also been used in the growth of large sapphire crystals[121], laser materials[201], and scintillators[202-206].

Spark plasma sintering is a technique that uses a large electric current in conjunction with applied compression pressure to transform and sinter materials. The electric current causes resistive (Joule) heating in the material, which provides the initial impetus for grain fusion and growth. The application of compressive pressure simultaneously forces individual grains closer together and increases both reactivity and enhances the sintering effect by increasing the amount of physical contact[207-208]. First patented in 1966[209], the technique can generate both single crystals and alloys and has become significantly more capable of producing materials for both research and for industrial use[122]. For example, spark plasma sintering has been used to synthesize several high entropy alloys[210-211], single crystals of $CoSb_3$[123] and SiC[212], important laser oxides such as YAG and Nd: YAG[213], $Bi_2S_3$[214], and uranium mononitride (UN)[215-216].

<u>Optical Floating Zone (OFZ) Growth – A Versatile Tool for Condensed Matter:</u>

The optical floating zone crystal growth technique utilizes focused light to melt a target material. Lamps, usually Xenon or Halogen, emit light that is then reflected and focused by large elliptical mirrors onto the target material. This growth technique is popular because it is broadly applicable to many different materials and the equipment is easily transitioned between different target materials. Additionally, the crystals produced by OFZ growth are of sizes (mm to cm) that are usable for many condensed matter characterization techniques. It also provides comparatively easy optical access for in situ monitoring and control.

<u>The Optical Floating Zone (OFZ) growth environment can be thought of as greenhouse effect:</u>

Similar to the greenhouse effect, when melting a material, we rely on the wavelength absorbed and consider the material to be a good black body. If we consider the sun's radiation is absorbed by the surface of the earth, then it is only less than half of what the sun radiates. Thankfully, there are various limiting factors that make the earth a non-ideal blackbody and the stratified mediums above the earth's good heat sink. Like the concept of the greenhouse effect, the OFZ utilizes some media that do not absorb any of the wavelengths of light such as the quartz chamber that shields the material and the synthesis gas environment. Whereas the material that needs to be in the molten state should absorb the optical light and thermally populate the energy from light into heat to melt the material.

Temperature is hard in Optical Floating Zone:

The optical floating zone method does not directly control temperature – instead, it controls the intensity and spectral distribution of light incident upon a specimen. We are trained to think that there is a direct relationship between input power and temperature from the Stefan–Boltzmann law, however, like many laws it has consideration. The consideration in the Stefan–Boltzmann law assumes that the object is a black body. Most materials are not perfect black bodies, and such considerations are limited since we do not have data on all the materials to calculate the exact temperature at a given OFZ power. Most useful synthesis databases such as the American Society of Metals (ASM) phase diagrams rely on composition as a function of temperature, however, we do not have databases that relate wavelength absorption as a function of composition[217]. This lack of correlation makes it challenging to design synthesis conditions – these same factors make pyrometery, a common non-contact temperature measure, difficult to apply quantitatively.

The wavelength dependence on heating is also the reason why the outer quartz doesn't melt on applying optical power. Typically, the melting point of quartz is around $T$~1600°C, however in practice the quartz softens and begins to crystallize at about $T$~1250°C (this is due to the water vapor that catalyzes bond rearrangements). Most materials such as iron or niobium have melting points below $T$~3600°C and are able to melt since they are good black body absorbers while quartz doesn't melt since it doesn't absorb visible or near IR optical light (it does absorb in the mid IR). This behavior alludes to the fact that optical floating zones depend on optical power absorption and not direct melting temperature. One might argue that the actual temperature can be measured in situ, e.g. through use of a pyrometer. In most cases, such a pyrometer reading is not accurate due to both unknown emissivities and contamination by stray light from the heating sources. However, one effective source to know the true temperature is to mount a thermocouple in the center of the rod during crystal growth. In this manner the temperatures near (but not in) the molten zone can be obtained[218]. However, engineering such thermocouple embedded powdered rods are challenging to make since sintering of such rods requires painstaking time and resources (and also precludes growth using the sections in which the thermocouples are embedded). The challenge of measuring the temperature of the molten zone (and its gradients) precisely is still a quite significant challenge.

In this respect, one recent advance is the development of laser OFZ techniques for single crystal growth[219-221]. The very nature of the floating zone furnace dictates that the molten zone will not be the only material in the furnace that gives off measurable light. Optical floating zones use mirrors to reflect the light given off by a lamp source, which by nature is omnidirectional. This means that there will be a significant amount of background light being given off by the lamp, in addition to light being reflected from the furnace mirrors onto the target material. This background light will be picked up by pyrometers and because those instruments typically do not have a way to distinguish light emitted from the molten zone from background and reflected light from other sources, the overall measure of temperature from the pyrometer becomes very unreliable. Laser diode floating zones do not suffer from this problem because instead of using mirrors to reflect omnidirectional light, focused laser light is used instead to heat the target material. The directional nature of lasers cuts down significantly on background light and increases the proportion of light being emitted from the molten zone. Furthermore, it is possible to position a furnace viewport and the pyrometer in such a way that it is impossible for ambient light to enter through the viewport to reach the detector, either directly or via reflection. This makes light pyrometry a more accurate measure of the molten zone temperature. However, if the reflectivity of the molten material increases relative to the solid, it is possible for laser light to be reflected off the melt and enter the pyrometer detector in that manner. As such, there is still a limit to how reliable the temperature reading of the melt is.

Related to the difficulty of precisely measuring the amount of light emitted by the molten zone is the need to know the emissivity of the target material and set the pyrometer correctly for this parameter. The emissivity of each material is different and is also different at different temperatures, making it extremely difficult to track the temperature accurately as the target material is heated. Even if the emissivity of the material is known at room temperature, often emissivity will not be known for the entire temperature range over which the material is being heated, which renders subsequent temperature measurements from the pyrometer unreliable.

The difficulties of obtaining a precise molten zone temperature present a potential opportunity for the use of alternative methods of temperature measurement. Many pyrometers measure the emission of infrared light as their primary means of calculating temperature. However, materials being heated emit both visible and infrared light. It is possible to take advantage of the visible spectrum radiation by using hyperspectral imaging, which produces a complete visible light spectrum at the point the camera is viewing in real time. Because the spectra are taken in real time, it is possible to distinguish the visible light being emitted from the molten zone from light reflected in the background, which improves the precision of the temperature measurement.

## Variations in common techniques

### Non-Intermetallic Melting via Arc Melting

There are many positives to arc melting synthesis techniques including reaching $T$~3500°C in a short duration, splat cooling for rapid quenching, the ability to rapidly alloy or purify large bulk materials in relatively short time, and casting rods or other desired morphologies while alloying. However, traditionally arc melting is done to alloy intermetallic materials under inert conditions such as Ar. The reason for utilizing intermetallics is because during arc melting, we want materials that can conduct the arc current, and metals are excellent in doing so. Most oxide materials are insulating and do not conduct the arc. However, simple proximity to an electric arc combined is often sufficient to melt the material anyway, and this is routinely employed in the growth of, e.g., cubic zirconia and magnesium oxide single crystals[222-223]. Another variation involves encapsulating the oxide and covering it with a surface metallic layer which arc conducts the arc, the heat of the motel metal will then transfer to the inner oxide capsule and thus melt oxide with the metal layer, thus alloying an oxide mixture. This method could be done with or without Ar environments with back filling of oxygen, nitrogen, and even atmospheric air in some cases based on the synthesis conditions. Such synthesis styles can also be incorporated for nitride synthesis under nitrogen gas in the arc melter. Sometimes adding drierite to arc melter, can allow to carry out synthesis under drying conditions which is sometimes useful when the humidity levels in the air are elevated. Thus, allowing us to explore the synthesis of novel oxides, nitrides, and intermetallics that require high melting points and rapid annealing.

Laser Pedestal in an OFZ

The laser pedestal technique was advanced to enable growth of very high melting oxides by requiring melting of only a small portion of the total material[224]. This functions similarly to the Czochralski technique by pulling out a single crystal from the melt on top of the "pedestal", and can also be performed in an OFZ furnace. Instead of generating a full molten zone connecting two solid rods, the tip of a single rod is heated to create a melt on the top of the rod. A small piece of wire (often metal) is then inserted into the melt and slowly drawn out of the melt, which crystallizes as the molten material leaves the hot zone. The melt is rotated at a constant speed counter to the rotation direction of the wire to stir the melt and produce a more homogeneous crystal. The slow draw speed out of the melt is facilitated by moving the wire and the rod in the same direction with the wire moving slightly faster[224-225]. This generates a Czochralski-like effect where, ideally, a single crystal is pulled out of the melt and cools slowly via the movement of the wire. This variant technique is effective for materials that can be melted but have low viscosity in the liquid state or other undesirable liquid properties that can prevent formation of a proper molten zone, but a few special considerations must be considered. The pulling wire material must be carefully selected such that it will not dissolve in the melt. Not only would this result in contamination of the entire melt and require it to be resolidified and removed, but also this situation leads to an automatic failure of the crystal growth. Additionally, the molten material must be viscous enough that it will adhere to the wire and be pulled for successful crystallization without detaching from the rest of the melt. If the melt is not viscous enough, it will not adhere to the wire and cannot be crystallized. If the melt is too viscous, it will be more difficult to maintain a stable melt because a significant amount of molten material will adhere to the wire, creating a large unstable bulge of molten material that can fall off the wire. Here the ability to tilt/angle the incident lasers becomes essential, because it permits use of a larger diameter "bottom" rod in which a pool of molten material can be produced with the material as its own crucible. This makes it a substantially easier to use version of a typical cold-crucible CZ/skull-melting process, but with similar ability to produce high purity and uniform materials, while being more energy efficient (since only a small portion of the material is melted at once).

Flux-like Traveling solvent technique for floating zone (TSFZ)

The traveling solvent (TS) technique is much like a flux crystal growth technique. In flux synthesis, the flux solvent is utilized to dissolve target phase and slowly cooled to crystallize, then the target phase crystals are separated from the flux solvent. In general, the flux technique is commonly utilized when the target phase is an incongruent melter, but is sometimes used for congruent melting phases to attain single crystals after dissolving the flux solvent[226-227]. The concept of traveling solvent in floating zone crystal growth is like the flux growth technique, in the sense that incongruent melting phase can be corrected via traveling solvent to attain phase purity of the target phase. In the TS method, the solvent is usually packed as a pellet or sintered at the end of the seed rod. The overall density of this TS pellet or region is lower than that of the feed rod. Therefore, the TS solution floats on top the molten zone ensuring that cooled crystallites settling on the seed rod are of the targeted composition. If the solvent pellet is used up or lost, the concentration of the melt may change, resulting in a compositional gradient along the grown crystal. This TS technique, much like the flux crystal growth technique, allows us to access phases that are incongruently melting which then allows us to synthesize challenging crystal systems. One example of a material grown using the TSFZ technique is GaAs, using elemental Ga as the traveling solvent. Extensive purification was seen and the use of the traveling solvent was useful for lowering the melting point of the target material for crystal growth[226].

Bridgman Technique in the Optical Floating Zone (OFZ)

Typically, the Bridgman technique is applied in an induction or resistive furnace or similar as it is relatively simple to generate a small, static hot zone with a well-defined temperature gradient, and oxygen-reactive materials can be easily grown using a sealed quartz tube to exclude atmosphere[228-229]. However, the Bridgman technique is also possible in the floating zone environment due to the ability to form a static heating zone. Instead of moving the seed and feed rods at different speeds, simply moving both rods at the same speed and in the same direction through the static heating area allows for Bridgman-type crystal growth and zone refinement. This variant technique is more effective when using a laser diode floating zone furnace compared to the traditional optical floating zone because of the greater precision of laser heating. Laser light can be focused on a smaller area much more easily than the light from the xenon lamps used in the optical floating zone; the latter typically requires several mirrors built specifically to reflect and focus the light from the xenon lamp, and the heating zone is generally larger as a result. Furthermore, this method can be used where a traditional Bridgman crystal growth might not be possible, perhaps due to sensitivity of the material to oxygen or moisture and would therefore need to be contained in a sealed container, or the material may be insufficiently conducting, rendering induction heating unsuitable. Optical and laser heating of a material is dependent only on its absorptivity, providing an alternative route towards melting and growing a single crystal material.

As noted before, in the optical floating zone technique, the absorption is limited to materials that absorb the laser light. During Bridgman growth, utilizing crucibles that do not react with the precursors becomes exceedingly important in targeting the desired phase. Boron nitride (BN) crucibles are virtually non-reactive to most materials; however, they are not good light absorbers of optical floating zone power which makes it challenging to conduct Bridgeman synthesis in a near IR laser floating zone. To address this concern, one can shade the outer region of the BN crucible with graphite, with light absorption depending on the coating of shading (darker shades will be better in light absorption – with the limit being the scenario of placing a BN crucible inside a C crucible).

Adding 0.001% of a color center dopant in transition metal to transparent materials acts effectively as an absorber and enables a variation of the traveling solvent floating zone (TSFZ) method where the color center (due to entropic considerations) preferentially stays in the melt, not in the growing crystal. In a crystal synthesis when a polycrystalline version of the target material is used as a precursor and the resulting crystal is of a different color than the polycrystalline powder, then the color change can be attributed to a difference in composition. A couple examples of such changes in color are in $Ba_2CaWO_{6-\delta}$[92] and $Yb_2Ti_2O_7$[93] single crystal syntheses. In $Ba_2CaWO_{6-\delta}$ the single crystal growth is done utilizing 7 bar pressure of Ar for stabilizing the growth. The resulting crystal had oxygen vacancies that were observed using the change in color from white to dark blue from polycrystalline to single crystalline forms[92]. To correct the change in color, the single was sintered under oxygen for a long duration of time to regain its white color.

Repeated Seeding in Crystal Growth

Oftentimes in the condensed matter community, the need of mm to cm scale single crystals becomes exceedingly important. Especially if one wants to study anisotropic thermal transport, neutron diffraction studies, or utilize single crystals as targets for growth of thin films amongst other uses. Growth of single crystals using the floating zone technique often yields large scale crystals of 2-15 cm scale. However, not all crystals can be grown via floating zone technique and sometimes utilizing techniques such as flux, chemical vapor transport, and hydrothermal synthesis is necessary. The crystals yielded in these techniques are usually small, of order 0.1-2 mm. Sometimes to address the size concern and to form larger crystals, seeded growths can be utilized. Here, taking small crystallites from initial growth attempts and reusing them with similar synthesis conditions can increase nucleation sites and increase the size of the crystal to double and sometimes triple the initial size of the crystal. Thus, creating avenues of formation of single crystals that meet the demands of size and scaling up.

An Open Frontier: Using Gradients in Crystal Growth

The use of temperature gradients in crystal growth is well-known and frequently used, most often while executing the Bridgman and Czochralski techniques. However, other parameter gradients can also be utilized to achieve the desired effect. These other parameters include, but are not limited to pressure, concentration, and magnetic and electric fields, all of which have been utilized in the preparation of "semiconductor" grade single crystals[85-90,96-97,230-233]. For example, concentration gradients can be seen in traditional dual solvent recrystallization techniques. The target material is dissolved in a solvent to the point of supersaturation, then a second solvent, in which the target material is both not present and insoluble, is carefully layered on top of the supersaturated solution. Over time, the two solvents mix, and the target material crystallizes out of solution. Concentration gradients are also used in flux crystal growths similarly to traditional solvent recrystallization, with the target material growing in areas of supersaturation within the flux[86,230-231]. The chemical vapor transport technique functions using a similar approach, but instead generating supersaturation via gradient in the gas phase, which allows for crystallization at one end of the tube where the supersaturation cannot be maintained, and the target material solidifies from the gas phase.

Non-standard gradients can be used to improve crystal growth by drawing impurities within the target material towards one end of the gradient[86]. This is most viable when there are significant differences in the physical properties of the target material and the known or suspected impurities, such as if the impurities are magnetic or electrically conductive and the target material is nonmagnetic or electrically insulating. For example, during an induction Bridgman growth, an electric current is induced in the target material which heats and eventually melts it. If there is an impurity phase which couples better than the target material with the applied current, it will remain in the melt while the rest crystallizes. Alternatively, it is possible to draw impurities to different areas of a solution using electric or magnetic fields, given that the impurity phase(s) and the target material respond differently, for example if the target material is nonmagnetic while the impurity phase is attracted to the field.

External applied pressure has been used in a number of different applications for material synthesis. Examples include synthetic diamonds, high temperature superconducting hydrides, and boron nitride[232-233]. Pressure gradients can have varying effects depending on the material. For example, AgI is shown to undergo a structural phase transition when shear stress is applied which then further transitions to different lattice types when additional pressure is applied[234]. Furthermore, it is also possible to dissociate AgI under applied pressure[235]. In terms of crystal growth, pressure gradients can be created as a result of the primary applied temperature gradient which then encourages crystal formation. For example, in physical vapor deposition and chemical vapor transport methods, a partial gas pressure composed of the gaseous target material or gaseous intermediate is generated at one end of the reaction chamber. This gas is immediately part of a pressure gradient across both ends of the chamber, following the applied temperature gradient. As the gas travels down the length of the chamber, the applied temperature gradient reduces the maximum vapor pressure of the gaseous material to the point where the maximum falls below the present vapor pressure at the cool end of the chamber, which causes crystal growth.

Summary: Several techniques for the synthesis of new materials, both common and uncommon, have been presented with a special focus on variations on the established methods as well as materials that are not typically synthesized with a specific technique. Important considerations prior to selecting some of the less well-known approaches for single crystal synthesis have also been discussed. Additional synthetic techniques in the scientist's toolkit provide improved flexibility,[236-237] for example, the ability to use the broader flux growth technique in the traveling solvent variant, which can help synthesize single crystals of materials that are incongruent melters, and the ability to target many more classes of materials for research. It is vitally important that the technique be understood so that variations can be utilized, especially if one wants to address common goals such as scaling up of single crystal growth. Other alterations of the techniques include carrying out synthesis under pressurized setups for the floating zone among others, seeded growth in chemical vapor transport, flux, and hydrothermal synthesis, and arc melting of non-intermetallic materials at high temperatures. With greater visibility, it is hoped, will come greater usage and throughput for new and interesting materials in order to push forward on the many challenges facing society today.

Acknowledgements: TB acknowledges support from NSF-MRSEC through the Princeton Center for Complex Materials NSF-DMR-2011750. TMM and NN acknowledge support by the National Science Foundation (Platform for the Accelerated Realization, Analysis, and Discovery of Interface Materials (PARADIM)) under Cooperative Agreement No. DMR-2039380.

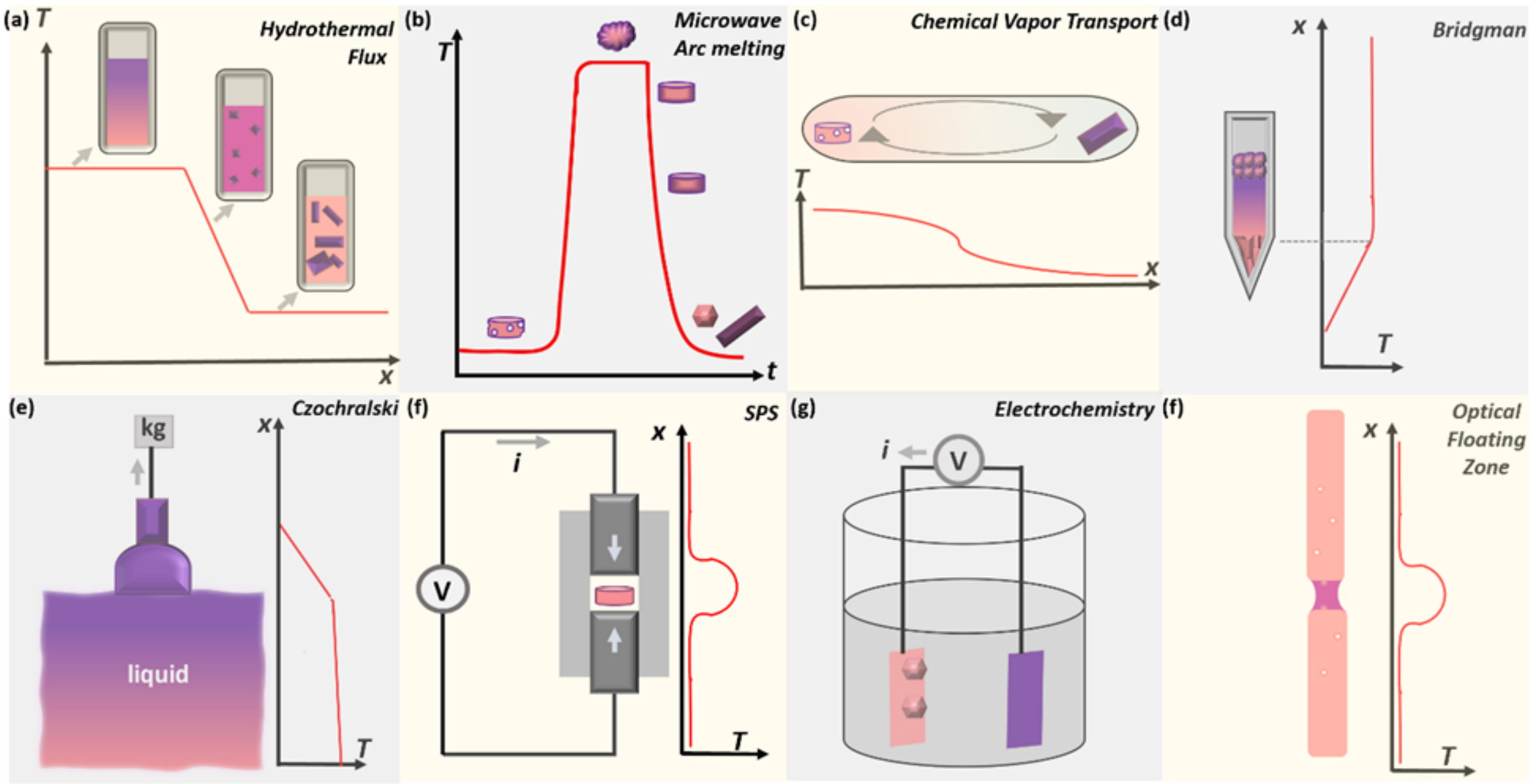

**Figure 1** Showcases a grid of synthesis techniques that are commonly utilized in solid state chemistry. More specifically **(a)** flux and hydrothermal synthesis, **(b)** microwave and microwave synthesis, **(c)** chemical vapor transport, **(d)** Bridgman, **(e)** Czochralski, **(f)** spark plasma sintering (SPS), **(g)** electrochemistry, and **(f)** optical floating zone methods are discussed above.

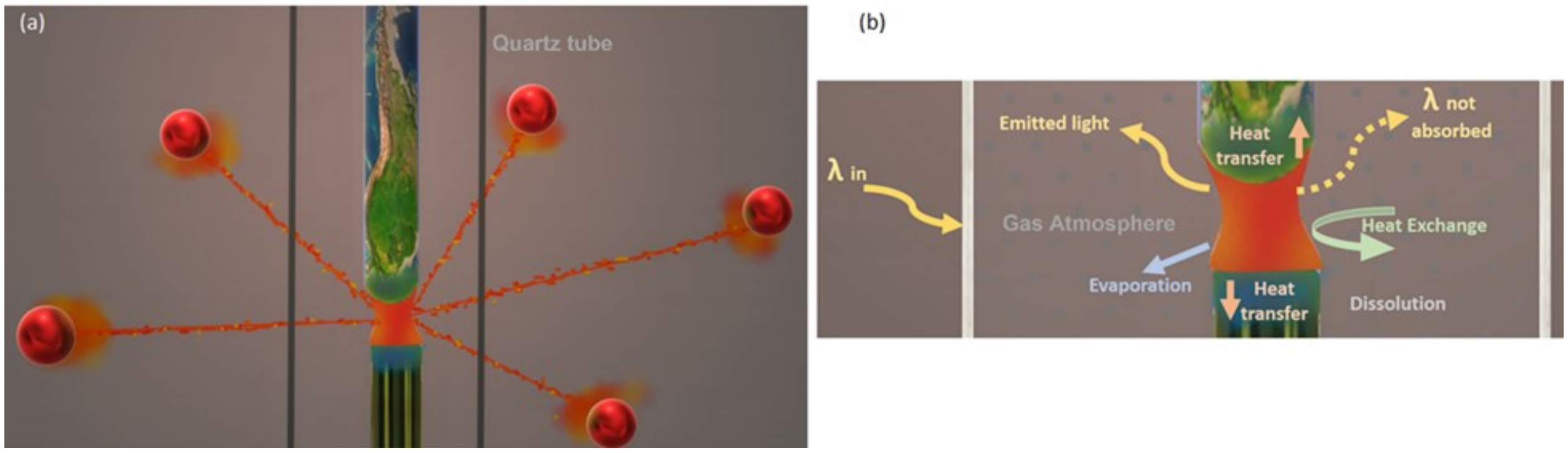

**Figure 2 (a)** Shows a traditional floating zone method that converts polycrystalline materials into phase pure single crystals. **(b)** This method can be thought of as the greenhouse effect that consists of wavelength of light that enters in provides heat source to create a molten zone, and avenues in which it undergoes heat sink and absorption.

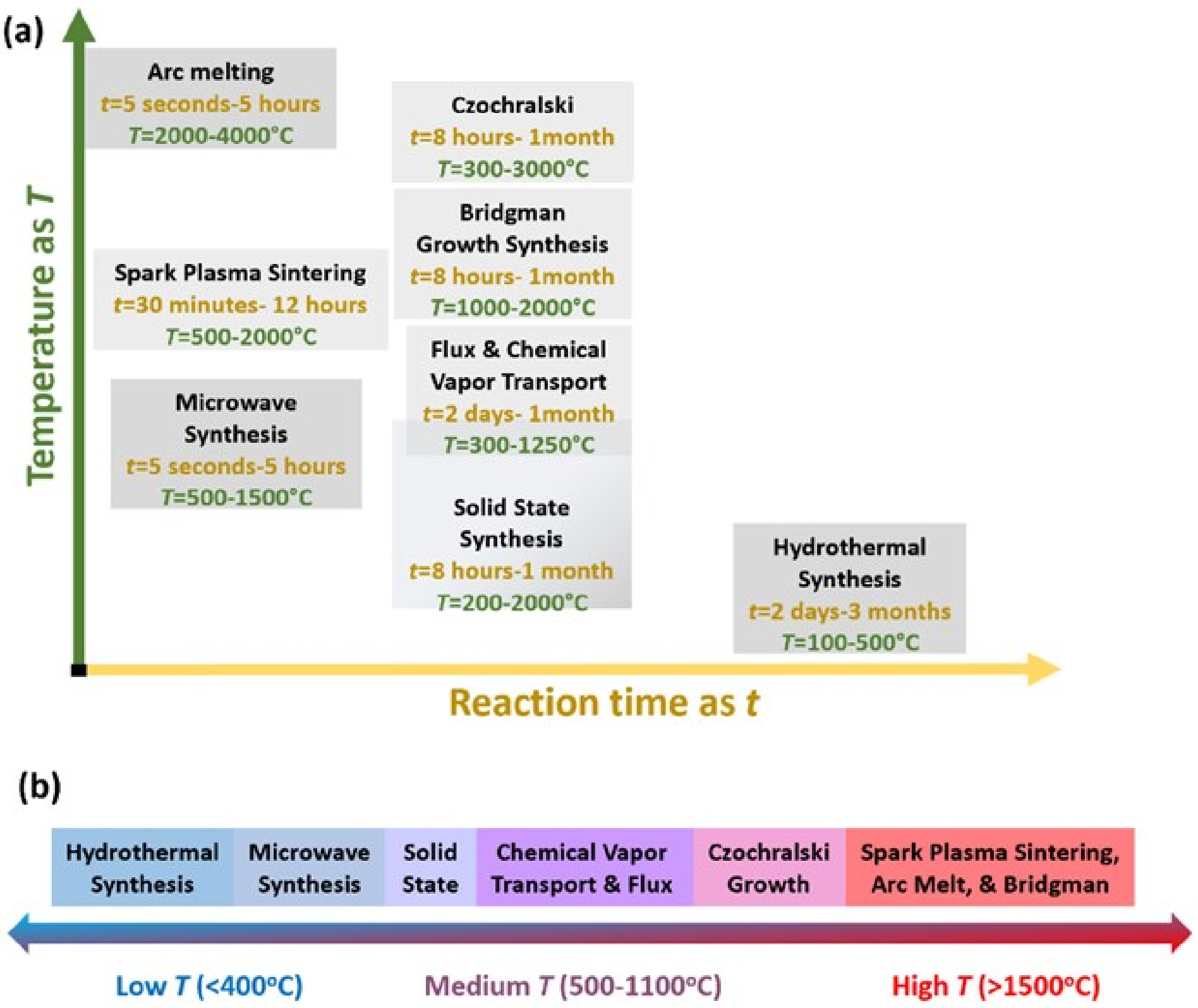

**Figure 3 (a)** Showcases a grid in which temperature as a function of reaction time is plotted for various single crystal synthesis techniques. **(b)** The temperature relationship with single crystal growth techniques from cold to hot.

| | **Oxides** | **Intermetallics** | **Nitrides** | **Sulfides** | **Halides** |
|---|---|---|---|---|---|
| **Floating Zone** | Rare earth, alkali, alkaline earth, & transition metal oxides[101] | RuAl, TiAl, TiAlNb, $Mn_3Si$, TiNb[102] | $Li_3N$[238]<br>NbN-NbC binaries[239]<br>$Cr_2N$[240]<br>TiN, ZrN[241] | CdS[242] | NaCl, KCl, KBr, KI, LiF[100] |
| **Flux Growth** | Transition metal, alkali, & alkaline earth oxides[105,135-138] | $AV_3Sb_5$, A = Alkali[61]<br>$EuGa_2Sb_2$[141] | h-BN[142]<br>GaN[143]<br>Multicomponent nitrides[243] | ZnS[139]<br>$NaCrS_2$, $NaInS_2$, CdS[140] | $Eu_4OCl_6$, $Eu_4OBr_6$[244]<br>$Tl_2MX_n$ (M = La, Hf; X = Cl, Br; n = 5, 6)[245] |
| **Chemical Vapor Transport** | ZnO[198], $WO_2$, $WO_3$[246] | CrB, $CrB_2$[246]<br>$Cr_3Si$, $Cr_5Si_3$, CrSi, $CrSi_2$[246]<br>$Ni_xGa_{1-x}$, $Cu_xGa_{1-x}$[247] | InN nanowires[248]<br>BN nanotubes[249] | $MoS_2$, $WS_2$[104,196], $ReS_2$, $Mo_2S_3$[196] | ZrNX (X = Cl, Br, I)[250-251]<br>CrOCl, $MoXY_n$ (X = O, S; Y = Cl, Br; n = 1, 2, 3), $WBr_2$[246] |
| **Arc Melting** | $CeMO_3$ (M = Al, Ga)[153]<br>$Zn_3Ta_2O_8$[132] | Rare earth intermetallics[133]<br>Transition metal intermetallics[134, 151, 252-253]<br>Medium- & high-entropy alloys[147-148] | BN nanotubes[154]<br>VN[254]<br>AlN nanowires & nanoparticles[255] | $Zr_{3+x}S_4$[155]<br>$Hf_2S$[256] | --- |
| **Bridgman** | $Y_3Al_5O_{12}$[63]<br>β-$Ga_2O_3$[257]<br>ZnO[258] | CdTe, CdZnTe[197] | --- | GaS[259]<br>$AgGaS_2$[260]<br>$Bi_2S_3$[261]<br>ZnS, CdS[262] | $KCaI_3$[263]<br>Alkaline earth mixed halides[264] |
| | | | | | Ternary alkali lead halides[265] |
| **Czochralski** | $BaTiO_3$, $TiO_2$[266]<br>β-$Ga_2O_3$[64,267]<br>$Y_3Al_5O_{12}$[64]<br>$Gd_3Ga_5O_{12}$[201]<br>$LiAlO_2$[268]<br>(La,Sr)(Al,Ta)$O_3$[269] | GaSb[270]<br>Rare earth tetraborides[271] | $Li_3N$[272] | ZnS, CdS[262] | Rare earth halides[273]<br>KCl[274]<br>BaBrCl[146]<br>$BaMgF_4$[64] |
| **Hydrothermal** | ZnO[124]<br>$WO_3$[176]<br>Transition metal oxides[125, 176-179] | $FeSn_2$[275]<br>IrRu[276]<br>$Pd_3Pb$[277]<br>SnSe[278] | VN[279]<br>BN[280]<br>GaN[174] | ZnS[124]<br>CdS[175]<br>$Co_2RuS_6$[281]<br>NiS, $Co_9S_8$[282] | $CsPb_2Cl_5$[283]<br>$CsPbBr_3$[173]<br>$Rb_2SeOCl_4$*$H_2O$[284]<br>$CsPb_2(Cl_{1-x}Br_x)_5$[283] |
| **Solvothermal** | Perovskite-structured oxides[285] | Binary Pt-based, Pd-based, and NiCo nanocrystals[286]<br>$Pt_2In_3$[287] | $Ta_3N_5$, TaN, MN (M = Zr, Hf, Nb)[288]<br>$Cu_3N$[289] | CdS[290]<br>$CdIn_2S_4$[291] | $CsPbX_3$ (X = Cl, Br, I)[292]<br>$CsSnX_3$ (X = Cl, Br, I)[293] |
| **Microwave** | $SnO_2$[294]<br>NiO[295]<br>Eu:$SrTiO_3$[180]<br>$LiV_3O_8$, $KNb_3O_8$, $KTiNbO_5$, $KSr_2Nb_3O_{10}$[112] | $Mg_2Sn$[296]<br>$Cu_{11}In_9$, $Ag_3In$, $AgIn_2$, $Ag_9In_4$, $AuIn_2$[297] | AlN[298-300], TiN, VN[299-300], GaN[300]<br>$Li_3FeN_2$, $Li_5TiN_3$, $Li_3AlN_2$[299] | ZnS nanoballs[159]<br>$Bi_2S_3$, $Sb_2S_3$[301]<br>ZnCdS[302]<br>$Ag_2S$, MS(M = Cd, Zn, Co, Pb, Cu)[303] | $CsPbX_3$ (X = Cl, Br, I)[304]<br>BiOX (X = Cl, Br, I)[305]<br>$Pb_5(VO_4)_3X$ (X = Cl, Br, I)[306] |
| **Spark Plasma Sintering** | Al:ZnO[307]<br>$Y_3Al_5O_{12}$[213] | $NbB_2$[309]<br>$CoSb_3$[123] | TiN[310]<br>UN[215-216] | $Cu_2S$[312]<br>$Bi_2S_3$[214] | $Tl_xCsBr$(X = 0.01, 0.1, |
| | $MoO_x$[308] | AlCuSiZnFe[149]<br>$AlFeCuCrMg_x$[210]<br>CoCrFeMnNi[211] | $Si_3N_4$[311] | $Fe_{7-x}Co_xS_8$[313]<br>$AgBi_3S_5$[314] | 0.2, 0.3, 0.5%)[315]<br>$Tl_xRbBr$(X = 0.1, 0.5, 1, 3%)[316] |

**Table 1.** A table showing a selection of different materials that can be crystallized by the following growth methods.

**<u>References</u>**